\documentclass[letter]{aa}  

\usepackage{subcaption}
\usepackage{lscape}         
\usepackage{placeins}

\usepackage{natbib}

\usepackage[english]{babel}
\usepackage[T1]{fontenc}
\usepackage[utf8]{inputenc}
\usepackage[normalem]{ulem}

\usepackage{mathtools}
\usepackage{physics}
\usepackage{xcolor}
\usepackage{graphicx}

\usepackage{placeins}
\usepackage{csquotes}

\usepackage{bm} 
\usepackage{balance}
\usepackage[pdftex, pdftitle={Article}, pdfauthor={Author}]{hyperref} 
\usepackage{stmaryrd}
\usepackage{tabulary}

\usepackage{txfonts}

\usepackage{orcidlink}

\renewcommand{\leq}{\leqslant}

\definecolor{Orange}{rgb}{1.0,0.5,0.15}
\definecolor{Blue}{rgb}{0,0.08,0.65}
\definecolor{Red}{rgb}{0.65,0.08,0.05}
\definecolor{Green}{rgb}{0.15,0.45,0.25}
\definecolor{Pink}{rgb}{1.0,0.05,0.5}
\definecolor{bubbles}{rgb}{0.91, 1.0, 1.0}
\definecolor{aquamarine}{rgb}{0.5, 1.0, 0.83}
\definecolor{bubblegum}{rgb}{0.99, 0.76, 0.8}
\definecolor{bluebell}{rgb}{0.74, 0.74, 0.92}
\definecolor{dollarbill}{rgb}{0.72, 0.93, 0.6}

\newcommand{\observed}[1]{\tilde{#1}}
\newcommand{\BNT}[1]{\hat{#1}}
\newcommand{\zb}[1]{z_b^{#1}}

\makeatletter
\renewcommand*\aa@subabstractiiname{Aim}
\makeatother

\begin{document}

   \title{Novel geometrical test of cosmological expansion from photometric data}

   \author{David~Touzeau\inst{1}\corrauth{david.touzeau@ipht.fr}        
        \and Francis~Bernardeau\inst{1}
        \and Karim~Benabed\inst{2}
        \and Sandrine~Codis\inst{3}
        }

   \institute{Universit\'e Paris-Saclay, IPhT, DRF-INP, UMR 3681, CEA, Orme des Merisiers Bat 774, 91191 Gif-sur-Yvette, France
   \and Sorbonne Universit\'e, CNRS, UMR 7095, Institut d'Astrophysique de Paris, 98 bis Boulevard Arago, 75014 Paris, France
   \and Universit\'e Paris-Saclay, Universit\'e Paris-Cit\'e, DAP, UMR 7158, CEA, Orme des Merisiers Bat 709, 91191 Gif-sur-Yvette, France
   }

   \date{Received \today}
 
    \abstract
   {}
   {In tomographic cosmic-shear observations, the Bernardeau, Nishimichi, Taruya (BNT) transform allows us to build weak lensing transformed maps for which the contribution from low-redshift lenses is nulled. As this transformation depends specifically on the expansion rate of the Universe and is independent of the matter distribution properties, it can be leveraged to extract information from large-scale structure probes at arbitrary non-linear scales. Thus, it provides constraints on cosmological background evolution.}
   {We demonstrate the applicability of the BNT transform as a probe by designing a null test and evaluating its performance for Stage IV weak lensing projects.}
   {Using a Fisher matrix analysis combined with parameter sampling, we show that this approach can significantly enhance constraints on the dark energy equation of state. In such surveys, apart from photometric redshifts uncertainties, performance is expected to be limited primarily by shape noise, rather than cosmic variance, making substantial improvements possible thanks to  better designed surveys in the future.}
   {}

   \maketitle
\nolinenumbers

\section{Introduction}
First described in \citep{Hu:1999ek}, the tomographic exploration of the cosmic shear signal has been made possible by the latest generation of large-scale galaxy surveys such as KiDS \citep{deJong:2012zb} \citep{KiDS:2020suj} and DES \citep{DES:2005dhi}, \citep{DES:2021wwk}. It is expected to reach new levels of performance with LSST \citep{LSSTDarkEnergyScience:2012kar} and Euclid \citep{Euclid:2024yrr}. These observations can be used to constrain cosmological parameters, based on the measured signal, which would be cosmic shear or galaxy number density, depending on the angular distances and on the amplitude of mass fluctuations.

Tomographic analyses make it possible to differentiate redshifts and therefore epochs of the expansion of the Universe, typically strengthening the constraints on the cosmological parameters. In such a construction, for a given bin of source redshifts, the cosmic-shear map results from the superposition of matter density fluctuations along the line of sight. The contribution of lenses is described by kernel functions, with their shapes given in Fig.~\ref{fig:BinsNulling}. 

In this context, the Bernardeau-Nishimishi-Taruya (BNT) transform, introduced in \cite{Bernardeau:2013rda}, is a technique used to build linear combinations of cosmic shear maps that are only sensitive to lenses from a restricted redshift range. Specifically, transformed cosmic shear maps can be made independent of low-redshift lenses. This property allows us to sort out the physical scales that contribute to the lensing signals at a given angular scale on the sky and are otherwise mixed by projection effects along the line of sight. This has practical applications, as described in \citep{Taylor:2020zcg,Bernardeau:2020jtc,Taylor:2020imc,Fronenberg:2023qtw,Maniyar:2021arp}. In this work, we rely on a different approach as we want to exploit the nulling property of the BNT transform and its dependence on the angular distance to directly constrain cosmological parameters.

Although this is not the first time the idea of exploiting a nulling property has been put forward \citep{Huterer:2005un,Joachimi:2010va,DES:2018lpj,DES:2021jzg}, however, this is the first time a full nulling property has been exploited outright \citep{Touzeau:2025lwu}. The idea we pursue here is to build a null test on well-chosen correlators to constrain the cosmological parameters involved in the distance-redshift relation. The goal is also to evaluate the constraining power of this approach, provided we can improve redshift precision, tomographic sharpness, and the level of shape noise.

\section{BNT nulling as a geometrical test}

More precisely, in this Letter, we illustrate a possible implementation of nulling and its performance in the context of a Euclid-like experiment using the setup described in \citep{Euclid:2019clj,Deshpande:2019sdl}. We refer to Appendix~\ref{sec:app} for details on the BNT nulling procedure (Eq.~\eqref{eq:BNT}) and the way we implement it in this Letter. On convergence maps, it reads as a linear transformation, $\BNT{\kappa}_a=p_{ai}\kappa_i$. It is clear that this transformation is independent of both scale and the regime of gravitational instabilities, as it only depends on the redshift distribution of the sources and the angular distance-redshift relation. This is particularly useful because it provides a new way to probe the expansion rate with weak lensing data, even on scales where simulations and modeling depend on poorly understood or unresolved small-scale effects. 

In practice, given a set of tomographic weak lensing observations and any other tomographic large-scale structure (LSS) tracer, the correlation between the LSS tracer at a specific redshift and the weak lensing observation (nulled by the BNT transform in that redshift range) should be zero, as long as the $p_{ai}$ coefficients are calculated for the correct cosmological model.
The constraints on cosmological parameters that can be obtained from such a nulling probe therefore rely entirely on how the coefficients $p_{ai}$ vary when the angular distance-redshift relation is changed. In particular, they are independent of both $\sigma_8$, the amplitude of the matter fluctuation, and $H_0$, two cosmological parameters that are affected by tensions between different measurements \citep{Abdalla:2022yfr}.

We note that constraints on $p_{ai}$ coefficients lead to specific constraints on the angular distance-redshift relation with some well-defined degeneracies between cosmological parameters. The identification of these degeneracies that we sketch in Appendix~\ref{sec:degen} is presented in detail in \citep{Touzeau:2025lwu}. It remains that nulling can effectively be used as a probe of the expansion rate of the Universe. The objective of this Letter is to assess the reach of such an approach.

\section{Forecasts for a Euclid-like setting}

In the remainder of this Letter, we investigate the efficiency of the nulling in a specific setting. As stated earlier, we follow the experimental setting of an Euclid-like survey. We further restricted ourselves to the case of a flat Universe, reducing the cosmological parameters to $\{\Omega_{\rm m},{\rm w}_0,{\rm w}_a\}$, which we set to the following fiducial values $\{0.32,-1,0\}$. 

The performance of the nulling probe ultimately relies on the significance with which we can assert that nulled convergence maps on one side and low-redshift LSS probes on the other are statistically independent. In this work, we studied one such null test where we use the observed cross-spectra $\observed{\mathcal{C}}^{\BNT{\kappa} {\rm g}}_{ai}(\ell)$ between nulled observed convergence maps, $\observed{\BNT{\kappa}}_a(\mathbf{n})$, and binned low-redshift galaxy density fields, $\observed{\delta}^{\rm g}_i(\mathbf{n})$. In the following, we denote the observed quantities as $\observed{X}$. We can compute this observable from the original cross-spectra, $\observed{\mathcal{C}}^{\kappa {\rm g}}_{ji}(\ell)$, between the original convergence maps $\observed{\kappa}_j(\mathbf{n})$, and binned low-redshift galaxy density fields, $\observed{\delta}^{\rm g}_i(\mathbf{n})$ as $\observed{\mathcal{C}}^{\BNT{\kappa} {\rm g}}_{ai}(\ell)=p_{aj}\observed{\mathcal{C}}^{\kappa {\rm g}}_{ji}(\ell)$. This is probably not the optimal strategy, but it has the advantage that its signal and noise can be fully evaluated from the data themselves as shown below, making the test fully independent on the details of the LSS statistical properties. 

In line with the procedure detailed in \citep{Touzeau:2025lwu}, we used low-redshift galaxy densities, regrouped into redshift bins, which are well chosen to optimize the coverage of low redshift space, while ensuring that these galaxies are indeed placed in the location where the transformed kernels are vanishing. Thus, we ensured there was no redshift overlap between low-redshift galaxy bins, $i$, and BNT-transformed kernels, $a$, involved in the computation of $\mathcal{C}^{\BNT{\kappa} {\rm g}}_{ai}(\ell)$. In our specific case, the BNT-transformed kernels were still organized per growing redshift of sources. The data vector can then be expressed as $\{\mathcal{C}^{\BNT{\kappa} {\rm g}}_{ai}(\ell)\}$ for $4 \leq a \leq 10$ and $1 \leq i \leq a-3$, with $ 32$ $\ell$-modes logarithmically spaced from $\ell=10$ to $\ell\approx16,000$.

At this point, we have the ability to build the null test. As we discuss above, we expect that when the cosmological parameters for the BNT transform have been aptly chosen, the expectation value of the power spectrum, $\observed{\mathcal{C}}^{\BNT{\kappa} {\rm g}}_{ai}(\ell)$, ought to be null. This is expressed as $\langle \observed{\mathcal{C}}^{\BNT{\kappa} {\rm g}}_{ai}(\ell) \rangle = 0$, for all $\ell$ and $i \leq a-3$. However, nulling is only exact when the elements of the deformation matrix, local shear, and convergence, are small. Any departures from this regime will break the linear relation structure in Eq.~\eqref{eq:convergence} and consequently lead to nulling. This nulling bias can occur at small angular scales when approaching a regime of strong lensing. Corrections at the bispectrum  level can then be used to infer the impact of such a bias. Two of its sources can be identified (see, for instance, \citep{Deshpande:2019sdl}): one from the so-called magnification bias and one from the reduced shear. All second-order corrections of the deformation matrix (lens-lens correlations and geodesic deviations) are reduced to $0$ assuming we are in the regime of validity of the Limber approximation. We note $\delta_B \mathcal{C}^{\BNT{\kappa} {\rm g}}_{ai}(\ell)$ this nulling bias correction that is proportional to the bispectrum and, thus, to the bispectrum normalization, $Q_3$, as well. Taking this bias into account, we have
\begin{equation}
        \langle \observed{\mathcal{C}}^{\BNT{\kappa} {\rm g}}_{ai}(\ell) \rangle = Q_3 \times \delta_B \mathcal{C}^{\BNT{\kappa} {\rm g}}_{ai}(\ell).
\end{equation}

This adds $Q_3$ as a new parameter to the system. Here, we set its fiducial value to 1 in our test data. Then, $\delta_B \mathcal{C}^{\BNT{\kappa} {\rm g}}_{ai}(\ell)$ is computed for fiducial parameters and used as a template. The introduction of this parameter as a nuisance parameter eventually helps define the angular scales that are not significantly affected by weak lensing corrections\footnote{The reduced shear and magnification bias corrections also lead to another bias involving the matter power spectrum. However, we do not need to correct for it since this bias is nulled for the same indices as $\observed{\mathcal{C}}^{\BNT{\kappa} {\rm g}}_{ai}(\ell)$.}. We assumed this approach would be enough to project the cosmological dependence of $\delta_B \mathcal{C}^{\BNT{\kappa} {\rm g}}_{ai}(\ell)$. We chose this one-parameter modeling procedure based on \citep{Scoccimarro:2000ee} and because it is sufficient in our simple proof-of-concept setup. This is also justified in the work by \citep{Touzeau:2025lwu}, where the calculation of the correction can be found. However, it is important to carefully validate this assumption or use a more realistic model in future applications to real data. We note that we do not include further discussions of a other observational systematics. In \citep{Touzeau:2025lwu}, the authors show that intrinsic alignments have no effect given our choice of observable. Based on the order of magnitudes given in \citep{Euclid:2024yrr}, multiplicative shear calibration bias is expected to be below the noise level for nulled correlators. Finally, photometric redshift uncertainties are discussed in \citep{Touzeau:2025lwu} and catastrophic outliers are left to a future work focused on a real setup.

Next, we can define a Gaussian likelihood $\mathcal{L}(\boldsymbol{\theta},Q_3)$ to jointly constrain the cosmological and $Q_3$ parameters via
{\small\begin{equation}
\label{eq:loglikelihood}
    \log\mathcal{L} = -\log(\mathcal{N}) 
    -\frac{1}{2} \sum_\ell (\observed{\mathcal{C}}^{\BNT{\kappa} {\rm g}}-Q_3 \delta_B \mathcal{C}^{\BNT{\kappa} {\rm g}})^T \cdot \Sigma^{-1} \cdot (\observed{\mathcal{C}}^{\BNT{\kappa} {\rm g}}-Q_3 \delta_B \mathcal{C}^{\BNT{\kappa} {\rm g}}),
\end{equation}}
where $\mathcal{N}$ is the normalization factor of the distribution, $\Sigma$ is the covariance, and we ignored the $\ell$ dependency and used a matrix notation avoiding $ai$ indices in the equation to make it more compact. We should recall that both the observed cross-spectra, $\observed{\mathcal{C}}^{\BNT{\kappa} {\rm g}}_{ai}(\ell)$, and the higher order bias correction, $\delta_B \mathcal{C}^{\BNT{\kappa} {\rm g}}_{ai}(\ell)$, as they are BNT-transformed quantities, depend on the cosmological parameters, $\boldsymbol{\theta}$.
We estimated the covariance, $\Sigma$, at leading order (i.e., ignoring correction from the Bispectrum term we discussed earlier) and assuming the classical Knox approximation \citep{Knox:1995dq}. In our setting, the only valid pair of indices, $a$ and $i$, are such that $1 \leq i < a-3 \leq 7$, leading to a $28 \times 28$ matrix:
\begin{equation}\label{eq:covmat}
\Sigma_{(ai),(bj)}(\ell)=\frac{\delta_{i,j}}{(2\ell+1)f_{\rm sky}} \left( \mathcal{C}^{\BNT{\kappa}\BNT{\kappa}}_{ab}(\ell)+\BNT{\mathcal{S}}_{ab}\right) \left( \mathcal{C}^{\rm gg}_{ii}(\ell)+\mathcal{N}^{\rm gg}_i \right).
\end{equation}
The covariance only contains the auto-spectra of the BNT transformed weak lensing maps,
$
    \mathcal{C}^{\BNT{\kappa}\BNT{\kappa}}_{ab}(\ell)=p_{ai}\mathcal{C}^{\kappa\kappa}_{ij}p_{bj},
$
and galaxy catalogs, $\mathcal{C}^{\rm gg}_{ii}(\ell)$, since all the cross-spectra between the two end up vanishing. The noise terms $\BNT{\mathcal{S}}_{ab}$ and $\mathcal{N}^{\rm gg}_i$ are, respectively, the BNT transformed shape noise, $\BNT{\mathcal{S}}_{ab}=p_{ai} \mathcal{S}_{ij} p_{bj}$, and the galaxy shot noise. Following \citep{Euclid:2019clj,Deshpande:2019sdl}, we set the shape noise at $\mathcal{S}_{ij}=\delta_{ij} \sigma_{\rm s}^2/\overline{n}_i$, with $\sigma_{\rm s}=0.3$ and the shot noise at $\mathcal{N}^{\rm gg}_{ij}=\delta_{ij}/\overline{n}^{\rm g}_i$, where $\overline{n}_i$ is the total number of galaxies (per $arcmin^2$) in bin $i$, with g indicating the low-redshift galaxy bins, and a sky fraction of $f_{\rm sky}=0.36$. We should note, as we discussed in \citep{Touzeau:2025lwu}, the covariance matrix is not affected by the super-sample covariance (SSC) nor by connected non-Gaussianities (cNG). As a result, the covariance matrix itself can be evaluated directly from measurements of the shear and galaxy clustering auto-spectra on the sky. At the price of a noisy covariance estimate, the implementation of the nulling test we propose is then entirely independent of small-scale theoretical modeling. This is also the case for the bispectrum level nulling bias correction template, $ \delta_B \mathcal{C}^{\BNT{\kappa} {\rm g}}_{ai}$.

We are now able to quantify the constraining power of these observables. Let us start with a simple Fisher matrix formalism to gain an initial understanding of their performance. The Fisher matrix is defined as
\begin{equation}\label{eq:FoM}
F^{\rm model}_{\theta \lambda} = \sum_{l,(ai),(bj)}  \Sigma^{-1}_{(ai),(bj)}(\ell) 
\frac{\partial  p_{ac}}{\partial \theta} \frac{\partial p_{bd} }{\partial \lambda}\mathcal{C}^{\BNT{\kappa} {\rm g}}_{ci}(\ell)\mathcal{C}^{\BNT{\kappa} {\rm g}}_{dj}(\ell),
\end{equation}
where $\theta$ and $\lambda$ denote the parameters of the model. These derivatives are computed analytically (i.e. with Mathematica). We note that here, Einstein's implicit summation is used on indices $c$ and $a$, while it is explicit for combinations $(ai)$ and $(bj)$. Correlators are computed under the Limber approximation. The inverse of $F_{\theta \lambda}$ is then the error covariance matrix of these parameters to first order in variations of the data vector. From this matrix, we get the figure-of-merit (FoM) in a flat Universe for the dark energy equation of state, which we define as the inverse determinant of the joint constraint on $\{{\rm w}_0,{\rm w}_a\}$ marginalized over the other parameters.
Unsurprisingly, nulling alone has a weak constraining power. However, it does allow us to derive cosmological constraints on the background evolution at scales that are beyond reach of standard $3\times2$pt analysis, that is to say, deeply into the nonlinear regime\footnote{This is a property shared with the approaches used in \citep{DES:2018lpj,DES:2021jzg} although the observable they use has to be calibrated with numerical simulations.}. 

\begin{figure}
  \begin{center}
    \includegraphics[width=0.9\columnwidth]{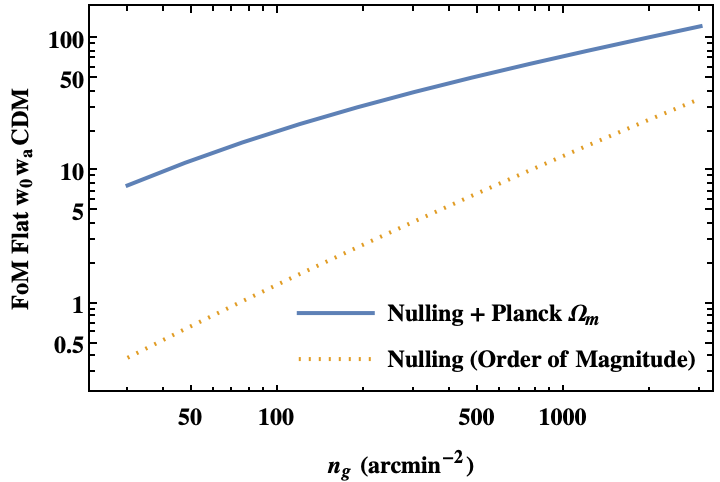}
  \end{center}
  \caption{Value of the FoM for the parameters ${\rm w}_0$, ${\rm w}_a$, obtained with nulling and with Planck's prior on $\Omega_{\rm m}$, as a function of the number density of sources to be used for shear measurements. We show for comparison a curve without Planck's prior on $\Omega_{\rm m}$, which is only an order of magnitude due to non-Gaussianity.}
  \label{fig:FoMtoNG}
\end{figure}

We illustrate in Fig.~\ref{fig:FoMtoNG} its constraining power when combined with CMB priors on $\Omega_{\rm m}$ derived from the Planck mission \citep{Planck:2018vyg}. We used the constraint for ${\rm w}_0{\rm w}_a$CDM cosmology from Planck and Baryon acoustic oscillation (BAO) data from BOSS DR12, MGS, and 6DF, which we multiplied by $2$ to be more conservative, which gives $\sigma(\Omega_{\rm m})=0.052$.
As we can see from Fig.~\ref{fig:FoMtoNG}, adding Planck's constraints on $\Omega_{\rm m}$ improves the FoM by one order of magnitude\footnote{When using nulling alone, the Fisher approach is not valid due to non-Gaussianity. Hence, this case, and the orange curve of Fig.~\ref{fig:FoMtoNG}, is to be taken as order of magnitude only.}. For a Euclid-like setting, the constraints are found to be modest. However, in Euclid-like settings, Fig. \ref{fig:FoMtoNG} also shows that one major limitation of this method is shape noise. Its constraining power would significantly improve if shape noise were reduced (e.g., by increasing the number of source galaxies) before ultimately being limited by nonlinear effects at small angular scales. This opens up the possibility, at least from a theoretical point of view, for future dedicated surveys to harness the statistical power of cosmological fluctuations across nearly all scales, even though accurate modeling at those scales remains beyond our current capabilities.
This stands in contrast to the standard approach, such as the use of 3×2pt correlation functions, which is fundamentally limited by our ability to model small-scale effects such as nonlinear clustering and baryonic physics. As a result, standard methods are effectively restricted to scales where shape noise is already relatively modest (see the Euclid Overview paper \citep{Euclid:2024yrr}) and rely on a number of theoretical priors to which the nulling method we propose is much less sensitive.

\begin{figure}
   \begin{center}
    \includegraphics[width=0.9\columnwidth]{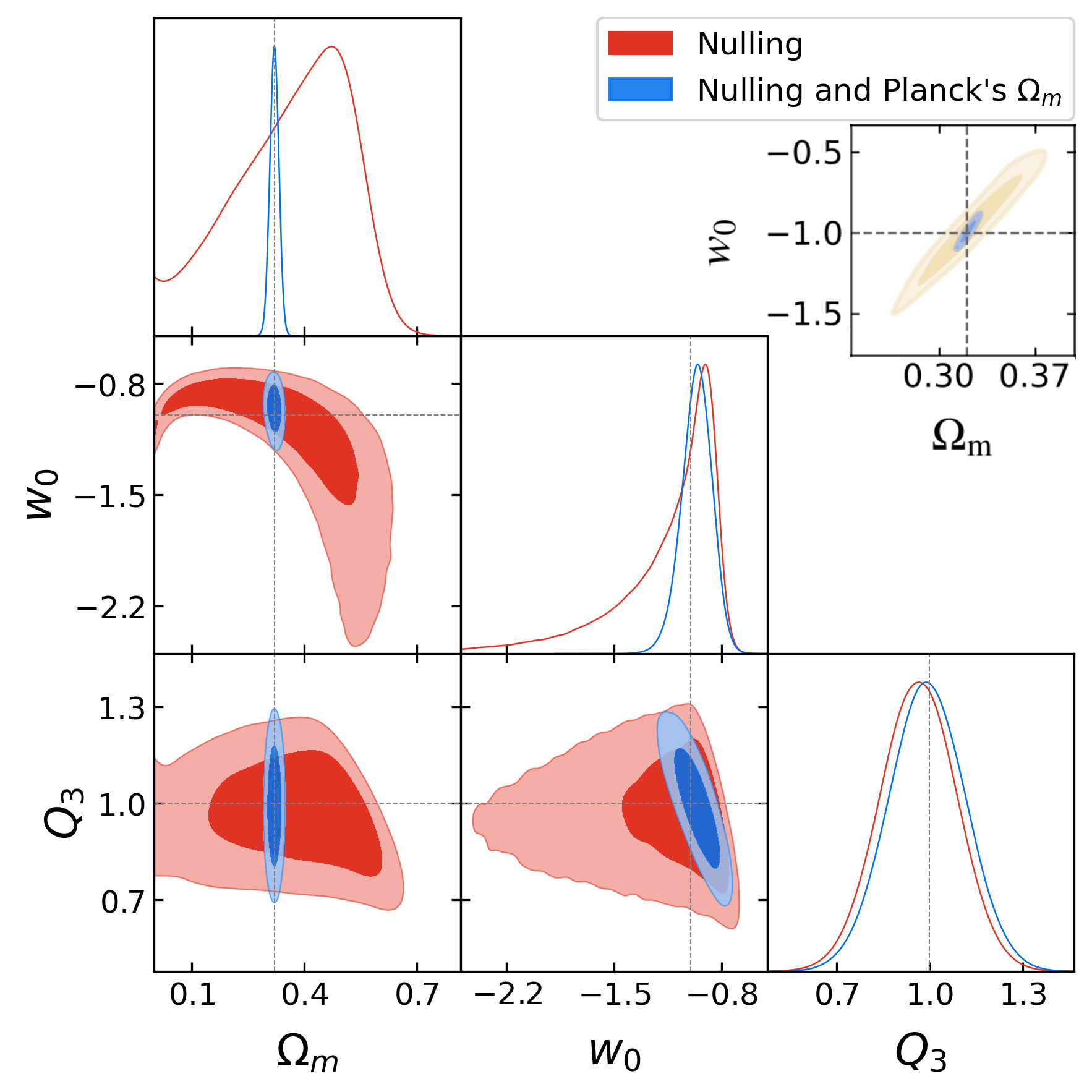}
  \end{center}
  \caption{Triangle plot for sampled parameters $\{\Omega_{\rm m},{\rm w}_0,Q_3\}$ with large uniform priors in blue and Planck+BAO w0-CDM constraint as a prior for $\Omega_{\rm m}$ in red. Dashed lines show the fiducial values of parameters. We show, in the top right corner, the Euclid forecast from the overview paper \citep{Euclid:2024yrr}.}
  \label{fig:MCMCpla}
\end{figure}

To gain further insights into the constraining power of this method, its sensitivity to parameters, and their correlations or degeneracies, we complemented 
the previous results with Markov chain Monte Carlo (MCMC) parameter samplings, restricting our analysis to the flat case. 
In practice, we used cobaya \citep{Torrado:2020dgo,2019ascl.soft10019T} with our likelihood from Eq.~\eqref{eq:loglikelihood} and synthetic idealized data that we theoretically computed for correlators in the fiducial model under Limber approximation.
The results are first illustrated in Fig.~\ref{fig:MCMCpla} with the triangle plot obtained for $\Omega_{\rm m}$ and ${\rm w}_0$ when sampling the three parameters $\{\Omega_{\rm m},{\rm w}_0,Q_3\}$ with flat large priors in one case; and in the other case, combined with CMB priors on $\Omega_{\rm m}$ derived from the Planck mission \citep{Planck:2018vyg}. We used the constraint for ${\rm w}_0$CDM cosmology from Planck and BAO data from BOSS DR12, MGS, and 6DF, which gives\footnote{Note that it is not the same value as for the Fisher Matrix in Fig.~\ref{fig:FoMtoNG} as now we work in the ${\rm w}_0$CDM case while it was the ${\rm w}_0{\rm w}_a$CDM before.} $\sigma(\Omega_{\rm m})=0.012$.
This plot shows the localization of the $\Omega_{\rm m}-{\rm w}_0$ degeneracy in the nulling-only case. This is actually inherited from the global degeneracy deriving from Eq.~\eqref{Eq:degeneracy} which is investigated more thoroughly in \citep{Touzeau:2025lwu} along with a discussion on the information content of the cosmic shear null test. We can actually see that this direction of the degeneracy is different from those usually encountered in weak-lensing observations (e.g \citep{Euclid:2024yrr}), the latter being strongly sensitive to the growth rate of fluctuations as well. Thus$-$and since this null test can exploit unused small scales and tests the weak lensing kernels, rather than the matter density fluctuations$-$it is a complementary probe to the usual weak lensing analysis, and they should be applied in combination. In contrast, the direction of degeneracy is pretty similar to the one encountered when using BAOs or supernovae, as can be seen in Fig. 5 of \citep{DESI:2024mwx}.

In Fig.~\ref{fig:MCMCpla}, we also show that the use of a strong prior on $\Omega_{\rm m}$ alone, as provided by CMB observations, can allow for the determination of the dark energy equation of state with significant precision.
Those results are summarized in Table~\ref{tab:confidence}, which gives the size of the posterior confidence interval at $1\sigma$, regardless of whether Planck priors are included or not. The posterior we find on ${\rm w}_0$, when combined with Planck, indicates an efficient constraining ability on the dark energy equation of state, especially considering the low numerical cost and simplicity of nulling as a probe.

\begin{table}[!t]
\caption{\label{tab:confidence}Precisions at $1 \sigma$ obtained from our samples compared to those obtained by Planck experiment.}
\begin{tabular}{|c|c|c|}
\hline
 & $\sigma(\Omega_{\rm m})$  & $\sigma({\rm w}_0)$ \\
\hline
Nulling & $0.15$ & $0.227$ \\
Nulling and Planck+BAO $\Omega_{\rm m}$ & $0.012$ & $0.095$ \\
Planck+BAO w0CDM & $0.012$ & $0.067$ \\
\hline
\end{tabular}
\tablefoot{When the distribution is symmetric, we show $\sigma$, otherwise we show half the confidence interval size.}
\end{table}

\section{Conclusion}

The scientific exploitation of Stage IV cosmic shear surveys is particularly challenging, as their full exploitation depends heavily on accurate modeling of nonlinear physics. Consequently, the resulting constraints on key cosmological parameters are bound to be intricately linked with those describing nonlinear structure growth (see, e.g., \citep{Euclid:2024yrr} for Euclid).

The method we propose here relies solely on the existence of density fluctuations and, thus, we exploit these fluctuations down to sub-megaparsec scales. It is a geometrical test that separates background evolution and structure growth modeling, enabling weak lensing surveys to harness statistical power at small physical scales (and at a negligible computational cost).

Photometric redshift uncertainties are a limiting factor. Specifically, catastrophic outliers generate residual correlations that can bias the inferred geometrical constraints. The mean redshift errors were investigated in \citep{Touzeau:2025lwu}, where the authors proposed using the cosmic shear null test to better constrain photometric redshifts. In addition, for Stage IV surveys, shape noise appears to be another primary limitation. Nevertheless, we show that when it is combined with CMB data, this method alone can already place meaningful constraints on the angular diameter distance–redshift relation and on fundamental cosmological parameters. At the very least, in conjunction with standard approaches, it can provide valuable consistency checks.

More interestingly, in contrast to the angular scale of BAOs (whose determination is strongly limited by cosmic variance), we demonstrate that a huge number of modes can potentially be exploited for this geometrical test. This requires surveys with significantly reduced shape noise, likely necessitating the development of a new generation of surveys employing novel observational strategies.
For example, weak lensing measurements through line intensity mapping \citep{Fronenberg:2023qtw,Maniyar:2021arp} could dramatically increase the number of small-scale structures over which shear can be measured. In addition, the use of kinematic lensing \citep{Huff:2013dha,S:2022htv} could significantly reduce shape noise, thereby allowing for a full exploitation of the potential of this method.

\begin{acknowledgements}
This work has received funding from the Centre National d’Etudes Spatiales for travel and hardware. It was carried out using the Infinity Cluster hosted by the Institut d’Astrophysique de Paris. We also acknowledge the Euclid consortium for organizing events and for providing opportunities to present this work at various stages, and we thank several of its members for useful discussions. SC warmly thanks H. Dupuy-Velu for her contributions to an earlier version of this work.
\end{acknowledgements}
\bibliographystyle{aa}
\bibliography{bibliography.bib}

@article{Bernardeau:2013rda,
    author = "Bernardeau, Francis and Nishimichi, Takahiro and Taruya, Atsushi",
    title = "{Cosmic shear full nulling: sorting out dynamics, geometry and systematics}",
    eprint = "1312.0430",
    archivePrefix = "arXiv",
    primaryClass = "astro-ph.CO",
    doi = "10.1093/mnras/stu1861",
    journal = "MNRAS",
    volume = "445",
    number = "2",
    pages = "1526--1537",
    year = "2014"
}

@article{Bernardeau:2020jtc,
    author = "Bernardeau, Francis and Nishimichi, Takahiro and Taruya, Atsushi",
    title = "{Observing Baryonic Acoustic Oscillations in tomographic cosmic shear surveys}",
    eprint = "2004.03201",
    archivePrefix = "arXiv",
    primaryClass = "astro-ph.CO",
    reportNumber = "YITP-20-15",
    month = "4",
    year = "2020",
    journal=""
}

@article{Taylor:2020zcg,
    author = "Taylor, Peter L. and Bernardeau, Francis and Huff, Eric",
    title = "{x-cut Cosmic Shear: Optimally Removing Sensitivity to Baryonic and Nonlinear Physics with an Application to the Dark Energy Survey Year 1 Shear Data}",
    eprint = "2007.00675",
    archivePrefix = "arXiv",
    primaryClass = "astro-ph.CO",
    doi = "10.1103/PhysRevD.103.043531",
    journal = "Phys. Rev. D",
    volume = "103",
    number = "4",
    pages = "043531",
    year = "2021"
}

@article{Fronenberg:2023qtw,
    author = "Fronenberg, Hannah and Maniyar, Abhishek S. and Pullen, Anthony R. and Liu, Adrian",
    title = "{Constraining cosmology with the CMB{\texttimes}line intensity mapping-nulling convergence}",
    eprint = "2309.06477",
    archivePrefix = "arXiv",
    primaryClass = "astro-ph.CO",
    doi = "10.1103/PhysRevD.109.123518",
    journal = "Phys. Rev. D",
    volume = "109",
    number = "12",
    pages = "123518",
    year = "2024"
}

@article{Maniyar:2021arp,
    author = "Maniyar, Abhishek S. and Schaan, Emmanuel and Pullen, Anthony R.",
    title = "{New probe of the high-redshift Universe: Nulling CMB lensing with interloper-free line intensity mapping pair lensing}",
    eprint = "2106.09005",
    archivePrefix = "arXiv",
    primaryClass = "astro-ph.CO",
    doi = "10.1103/PhysRevD.105.083509",
    journal = "Phys. Rev. D",
    volume = "105",
    number = "8",
    pages = "083509",
    year = "2022"
}

@article{Deshpande:2019sdl,
    author = "Deshpande, A. C. and Kitching, T. D. and Cardone, V. F. and others",
    title = "{Euclid: The reduced shear approximation and magnification bias for Stage IV cosmic shear experiments}",
    eprint = "1912.07326",
    archivePrefix = "arXiv",
    primaryClass = "astro-ph.CO",
    doi = "10.1051/0004-6361/201937323",
    journal = "A\&A",
    volume = "636",
    pages = "A95",
    year = "2020"
}

@article{Euclid:2019clj,
    author = "Blanchard, A. and Camera, S. and Carbone, C. and others",
    collaboration = "Euclid",
    title = "{Euclid preparation. VII. Forecast validation for Euclid cosmological probes}",
    eprint = "1910.09273",
    archivePrefix = "arXiv",
    primaryClass = "astro-ph.CO",
    doi = "10.1051/0004-6361/202038071",
    journal = "A\&A",
    volume = "642",
    pages = "A191",
    year = "2020"
}

@article{Scoccimarro:2000ee,
    author = "Scoccimarro, Roman and Couchman, H. M. P.",
    title = "{A fitting formula for the nonlinear evolution of the bispectrum}",
    eprint = "astro-ph/0009427",
    archivePrefix = "arXiv",
    doi = "10.1046/j.1365-8711.2001.04281.x",
    journal = "MNRAS",
    volume = "325",
    pages = "1312",
    year = "2001"
}

@article{Torrado:2020dgo,
    author = "Torrado, Jesus and Lewis, Antony",
    title = "{Cobaya: Code for Bayesian Analysis of hierarchical physical models}",
    eprint = "2005.05290",
    archivePrefix = "arXiv",
    primaryClass = "astro-ph.IM",
    reportNumber = "TTK-20-15",
    doi = "10.1088/1475-7516/2021/05/057",
    journal = "JCAP",
    volume = "05",
    pages = "057",
    year = "2021"
}

@software{2019ascl.soft10019T,
       author = {{Torrado}, Jes{\'u}s and {Lewis}, Antony},
        title = "{Cobaya: Bayesian analysis in cosmology}",
 howpublished = {Astrophysics Source Code Library, record ascl:1910.019},
         year = 2019,
        month = oct,
          eid = {ascl:1910.019},
       adsurl = {https://ui.adsabs.harvard.edu/abs/2019ascl.soft10019T}
}

@article{deJong:2012zb,
    author = "de Jong, Jelte T. A. and Verdoes Kleijn, Gijs A. and Kuijken, Konrad H. and Valentijn, Edwin A.",
    collaboration = "Astro-WISE, KiDS",
    title = "{The Kilo-Degree Survey}",
    eprint = "1206.1254",
    archivePrefix = "arXiv",
    primaryClass = "astro-ph.CO",
    doi = "10.1007/s10686-012-9306-1",
    journal = "Exper. Astron.",
    volume = "35",
    pages = "25--44",
    year = "2013"
}

@article{KiDS:2020suj,
    author = "Asgari, M. and Lin, C.-A. and Joachimi, B. and others",
    collaboration = "KiDS",
    title = "{KiDS-1000 Cosmology: Cosmic shear constraints and comparison between two point statistics}",
    eprint = "2007.15633",
    archivePrefix = "arXiv",
    primaryClass = "astro-ph.CO",
    doi = "10.1051/0004-6361/202039070",
    journal = "A\&A",
    volume = "645",
    pages = "A104",
    year = "2021"
}

@article{LSSTDarkEnergyScience:2012kar,
    author = {Abate, A. and Ansari, R. and Campagne, Jean-Eric and Moniez, M. and Perdereau, O. and Plaszczynski, St{\'e}phane and Tristram, M. and Boutigny, D. and Antilogus, P. and Astier, P. and Bongard, S. and Guy, J. and Pain, R. and Russo, S. and Ricol, J.S. and Ealet, A. and Escoffier, S. and Fouchez, D. and Tao, C. and Tilquin, A.},
    collaboration = "LSST Dark Energy Science",
    title = "{Large Synoptic Survey Telescope: Dark Energy Science Collaboration}",
    eprint = "1211.0310",
    archivePrefix = "arXiv",
    primaryClass = "astro-ph.CO",
    reportNumber = "FERMILAB-FN-0952-A-T",
    month = "11",
    year = "2012",
    journal=""
}

@article{DES:2005dhi,
    author = "Abbott, T. M. C. and Aldering, G. and Annis, James T. and others",
    collaboration = "DES",
    title = "{The Dark Energy Survey}",
    eprint = "astro-ph/0510346",
    archivePrefix = "arXiv",
    reportNumber = "FERMILAB-PUB-05-656-A",
    month = "10",
    year = "2005",
    journal=""
}

@article{DES:2021wwk,
    author = "Abbott, T. M. C. and Aguena, M. and Alarcon, A. and others",
    collaboration = "DES",
    title = "{Dark Energy Survey Year 3 results: Cosmological constraints from galaxy clustering and weak lensing}",
    eprint = "2105.13549",
    archivePrefix = "arXiv",
    primaryClass = "astro-ph.CO",
    reportNumber = "FERMILAB-PUB-21-221-AE, DES-2020-0617",
    doi = "10.1103/PhysRevD.105.023520",
    journal = "Phys. Rev. D",
    volume = "105",
    number = "2",
    pages = "023520",
    year = "2022"
}

@article{DES:2021jzg,
    author = "S{\'a}nchez, Carles and Prat, J. and Zacharegkas, G. and others",
    collaboration = "DES",
    title = "{Dark Energy Survey Year 3 results: Exploiting small-scale information with lensing shear ratios}",
    eprint = "2105.13542",
    archivePrefix = "arXiv",
    primaryClass = "astro-ph.CO",
    reportNumber = "FERMILAB-PUB-21-247-AE, DES-2020-0596",
    doi = "10.1103/PhysRevD.105.083529",
    journal = "Phys. Rev. D",
    volume = "105",
    number = "8",
    pages = "083529",
    year = "2022"
}

@article{DES:2018lpj,
    author = "Prat, J. and Baxter, E.J. and Shin, T. and others",
    collaboration = "DES, SPT",
    title = "{Cosmological lensing ratios with DES Y1, SPT and Planck}",
    eprint = "1810.02212",
    archivePrefix = "arXiv",
    primaryClass = "astro-ph.CO",
    reportNumber = "FERMILAB-PUB-18-547-AE",
    doi = "10.1093/mnras/stz1309",
    journal = "MNRAS",
    volume = "487",
    number = "1",
    pages = "1363--1379",
    year = "2019"
}

@article{Huterer:2005un,
    author = "Huterer, Dragan and White, Martin J.",
    title = "{Nulling tomography with weak gravitational lensing}",
    eprint = "astro-ph/0501451",
    archivePrefix = "arXiv",
    doi = "10.1103/PhysRevD.72.043002",
    journal = "Phys. Rev. D",
    volume = "72",
    pages = "043002",
    year = "2005"
}

@article{Joachimi:2010va,
    author = "Joachimi, B. and Schneider, P.",
    title = "{Controlling intrinsic alignments in weak lensing statistics: The nulling and boosting techniques}",
    eprint = "1009.2024",
    archivePrefix = "arXiv",
    primaryClass = "astro-ph.CO",
    month = "9",
    year = "2010",
    journal=""
}

@article{Planck:2018vyg,
    author = "Aghanim, N. and Akrami, Yashar and Ashdown, M. and others",
    collaboration = "Planck",
    title = "{Planck 2018 results. VI. Cosmological parameters}",
    eprint = "1807.06209",
    archivePrefix = "arXiv",
    primaryClass = "astro-ph.CO",
    doi = "10.1051/0004-6361/201833910",
    journal = "A\&A",
    volume = "641",
    pages = "A6",
    year = "2020"
}

@article{Hu:1999ek,
    author = "Hu, Wayne",
    title = "{Power spectrum tomography with weak lensing}",
    eprint = "astro-ph/9904153",
    archivePrefix = "arXiv",
    doi = "10.1086/312210",
    journal = "ApJL",
    volume = "522",
    pages = "L21--L24",
    year = "1999"
}

@article{Euclid:2024yrr,
    author = "Mellier, Y. and Uf, Abdurro and Acevedo Barroso, J.A. and others",
    collaboration = "Euclid",
    title = "{Euclid. I. Overview of the Euclid mission}",
    eprint = "2405.13491",
    archivePrefix = "arXiv",
    primaryClass = "astro-ph.CO",
    doi = "10.1051/0004-6361/202450810",
    journal = "A\&A",
    volume = "697",
    pages = "A1",
    year = "2025"
}

@article{Touzeau:2025lwu,
    author = "Touzeau, David and Bernardeau, Francis and Benabed, Karim and Codis, Sandrine",
    title = "{Cosmic shear nulling as a geometrical cosmological probe: Methodology and sensitivity to cosmological parameters and systematics}",
    eprint = "2502.02246",
    archivePrefix = "arXiv",
    primaryClass = "astro-ph.CO",
    doi = "10.1103/zcsl-4644",
    journal = "Phys. Rev. D",
    volume = "113",
    number = "12",
    pages = "123519",
    year = "2026"
}

@article{Taylor:2020imc,
    author = "Taylor, P.L. and Kitching, T. and Cardone, V.F. and others",
    title = "{Euclid: Forecasts for $k$-cut $3 \times 2$ Point Statistics}",
    eprint = "2012.04672",
    archivePrefix = "arXiv",
    primaryClass = "astro-ph.CO",
    doi = "10.21105/astro.2012.04672",
    journal = "Open J. Astrophys.",
    volume = "4",
    number = "1",
    pages = "6",
    year = "2021"
}

@article{Knox:1995dq,
    author = "Knox, LLoyd",
    title = "{Determination of inflationary observables by cosmic microwave background anisotropy experiments}",
    eprint = "astro-ph/9504054",
    archivePrefix = "arXiv",
    reportNumber = "FERMILAB-PUB-95-008-A",
    doi = "10.1103/PhysRevD.52.4307",
    journal = "Phys. Rev. D",
    volume = "52",
    pages = "4307--4318",
    year = "1995"
}

@article{Abdalla:2022yfr,
    author = "Abdalla, Elcio and Franco Abell{\'a}n, Guillermo and Aboubrahim, Amin and others",
    title = "{Cosmology intertwined: A review of the particle physics, astrophysics, and cosmology associated with the cosmological tensions and anomalies}",
    eprint = "2203.06142",
    archivePrefix = "arXiv",
    primaryClass = "astro-ph.CO",
    reportNumber = "FERMILAB-CONF-22-192-SCD",
    doi = "10.1016/j.jheap.2022.04.002",
    journal = "JHEAp",
    volume = "34",
    pages = "49--211",
    year = "2022"
}

@article{Huff:2013dha,
    author = "Huff, Eric M. and Krause, Elisabeth and Eifler, Tim and Fang, Xiao and George, Matthew R. and Schlegel, David",
    title = "{Cosmic shear without shape noise}",
    eprint = "1311.1489",
    archivePrefix = "arXiv",
    primaryClass = "astro-ph.CO",
    month = "11",
    year = "2013",
    journal=""
}

@article{S:2022htv,
    author = "Pranjal, R. S. and Krause, Elisabeth and Huang, Hung-Jin and Huff, Eric and Xu, Jiachuan and Eifler, Tim and Everett, Spencer",
    title = "{Kinematic lensing inference {\textendash} I. Characterizing shape noise with simulated analyses}",
    eprint = "2209.11811",
    archivePrefix = "arXiv",
    primaryClass = "astro-ph.GA",
    doi = "10.1093/mnras/stad2014",
    journal = "MNRAS",
    volume = "524",
    number = "3",
    pages = "3324--3334",
    year = "2023"
}

@article{DESI:2024mwx,
    author = "Adame, A.G and Aguilar, J and Ahlen, S and others",
    collaboration = "DESI",
    title = "{DESI 2024 VI: cosmological constraints from the measurements of baryon acoustic oscillations}",
    eprint = "2404.03002",
    archivePrefix = "arXiv",
    primaryClass = "astro-ph.CO",
    reportNumber = "FERMILAB-PUB-24-0154-PPD",
    doi = "10.1088/1475-7516/2025/02/021",
    journal = "JCAP",
    volume = "02",
    pages = "021",
    year = "2025"
}

\appendix
\section{Binning and nulling strategy}\label{sec:app}

\begin{figure}[!ht]
  \begin{center}
    \includegraphics[width=0.9\columnwidth]{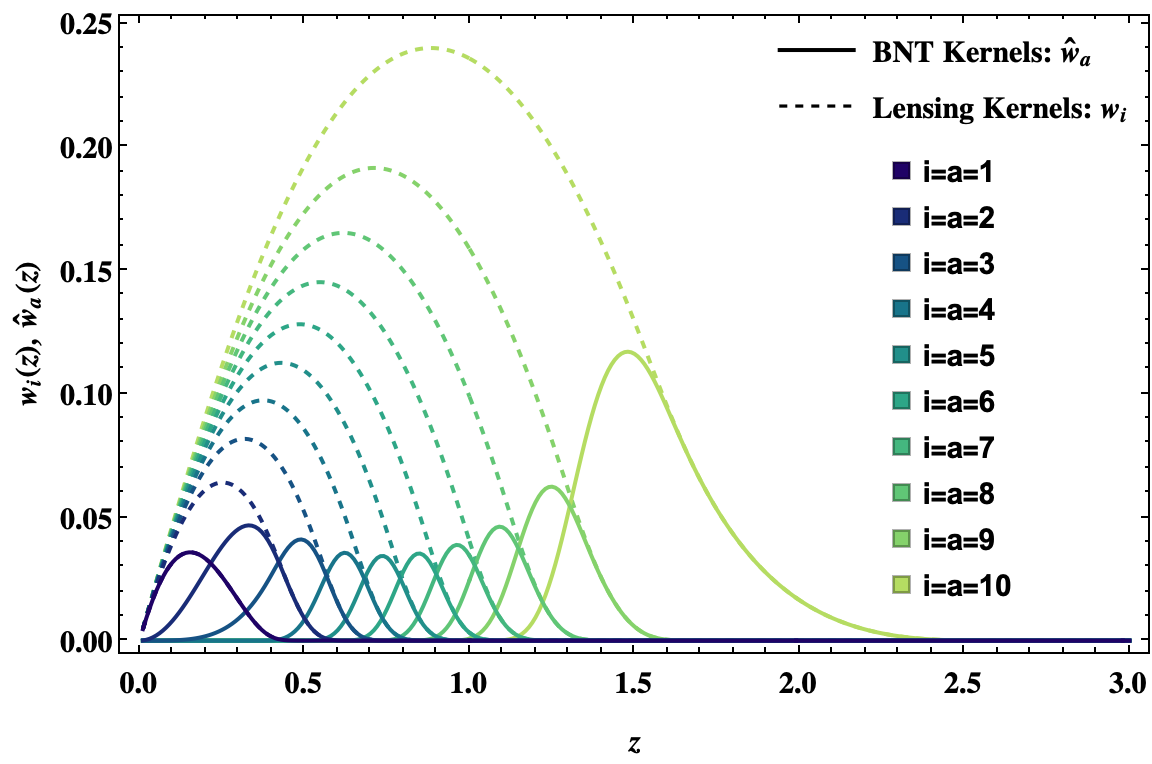}
  \end{center}
  \caption{Shape of the nulled lensing kernels $\BNT{w}_a(z)$ (solid lines) constructed for ten equally populated bins and of original lensing kernels ${\rm w}_i(z)$ (dashed lines).}
  \label{fig:BinsNulling}
\end{figure}

As in \citep{Euclid:2019clj}, we assume that we have observed positions and shears of a catalog of galaxies whose density distribution follows
\begin{equation}
    n(z)\propto \left(\frac{z}{z_0}\right)^2 \exp \left[-\left(\frac{z}{z_0}\right)^\frac{3}{2}\right],
\end{equation}
normalized for a galaxy density of $30 {\rm \ arcmin^{-2}}$, where $z_0=z_{\rm m}/\sqrt{2}$ and $z_{\rm m}=0.9$ is the median redshift of the survey. This distribution allows us to define $10$ equally populated photometric redshift bins\footnote{Taking into account realistic uncertainties in redshift determinations, the distribution of the true redshift of the galaxies selected in each of such bins depends on $p_{\rm ph}(z_p,z,\zb{i},\sigma_z)\,\dd z_p$, the probability that a galaxy of redshift $z$ has a measured photometric redshift equal to $z_p$ within $\dd z_p$. In the present study, it is  parameterized by a simple Gaussian distribution of width $\sigma_z$ and mean $\zb{i}$.} $n_i(z_s)$. Those bins define $10$ lensing efficiencies or lensing kernels, denoted ${\rm w}_i$ for each redshift bin $i$, with which convergence maps $\kappa_i$ can be reconstructed. In an idealized setting, these maps probe the large-scale structure along the line of sight,
\begin{equation}\label{eq:convergence}
    \kappa_i (\mathbf{n}) = \int_0^\infty \dd\chi {\rm w}_i(\chi) \delta_{\rm m}(\chi,\mathbf{n}),
\end{equation}
where $\mathbf{n}$ is the position of the source galaxy in the sky and the lensing efficiency in bin $i$ is computed from the redshift galaxy distribution $n_i(z)$ as ${\rm w}_i(\chi)=\int \dd z_{\rm S}\, n_i(z_{\rm S}) w(\chi(z_{\rm S}),\chi)$ with the usual lensing kernel,
\begin{equation}\label{eq:kernel}
    w(\chi_{\rm S},\chi)=\frac{3 \Omega_{\rm m} H_0^2}{2 c^2} \frac{D_K(\chi_{\rm S}-\chi)D_K(\chi)}{D_K(\chi_{\rm S}) a(\chi)} \Theta(\chi_{\rm S}-\chi),
\end{equation}
where $H_0$ is the Hubble constant, $c$ the speed of light, $\Omega_{\rm m}$ the matter content of the Universe at redshift $0$, while $\Theta$ is the Heaviside step function (noting there is no lensing effect from matter behind the source) and $D_K$ is the comoving distance, 
\begin{equation}
D_K(\chi)=
	\begin{cases}
	\frac{c}{H_0} \frac{\sin(\sqrt{K}\chi H_0/c)}{\sqrt{K}}  &$for $ K>0, \\
	\chi &$for $ K=0, \\
	\frac{c}{H_0} \frac{\sinh(\sqrt{-K}\chi H_0/c)}{\sqrt{-K}} \ \ \ &$for $ K<0,
	\end{cases},
 \end{equation}
where $K$ is the curvature. The lensing kernels, ${\rm w}_i$, are shown in Fig.~\ref{fig:BinsNulling}.

As shown in \citep{Bernardeau:2013rda}, the BNT transform is given as a matrix, $p_{ai}$, that transforms equivalently lensing kernels and convergence maps as 
\begin{equation}\label{eq:BNT}
    \BNT{w}_a=p_{ai}{\rm w}_i \ {\rm and} \ \BNT{\kappa}_a=p_{ai}\kappa_i.
\end{equation} 
Here, we used implicit summation on repeated indices. As illustrated in Fig.~\ref{fig:BinsNulling}, these transformed or "nulled" lensing kernels are more localized in redshift than the original. Thus, they overlap less, so that some maps are no longer correlated, while the remaining non-vanishing correlations are smaller.
The transformation is built from the two following quantities or "moments" via
\begin{equation}\label{eq:moments}
n_i^{(0)}=\int \dd\chi\, n_i(\chi) \ \ {\rm and} \ \ n_i^{(1)}=\int \dd\chi \,\frac{n_i(\chi)}{F_K(\chi)},
\end{equation}
where we define\footnote{In the case of zero-curvature, $D_K$ and $F_K$ are identical. See \citep{Bernardeau:2013rda} for details of why $F_K$ enters the expression of $p_{ai}$ in general.}
\begin{equation}
F_K(\chi)=
	\begin{cases}\frac{c}{H_0}
	  \frac{\tan(\sqrt{K}\chi H_0/c)}{\sqrt{K}}  &$for $ K>0, \\
	\chi
 &$for $ K=0, \\
	\frac{c}{H_0}\frac{\tanh(\sqrt{-K}\chi H_0/c)}{\sqrt{-K}} \ \ \ &$for $ K<0.
	\end{cases}
\end{equation}
Then the conditions to define the BNT transform matrix are
\begin{equation}\label{eq:BNTMatrix}
\sum_{i=a-2}^a p_{ai} n_i^{(0)}=0 \ {\rm and} \ \sum_{i=a-2}^a p_{ai} n_i^{(1)}=0.
\end{equation}
The coefficients are normalized with $p_{ii}=1$. As the first two lines of the matrix have less non vanishing coefficients, we impose $(1,0,...,0)$ for the first one and $(-1,1,0,...,0)$ for the second one. The rest of the matrix is defined by Eq.~\eqref{eq:BNTMatrix}. This translates into a lower triangular band matrix whose values under the second lower diagonal are null.

\section{Degeneracy}\label{sec:degen}

In this section, we sketch the derivation of the degeneracy relation presented in detail in \cite{Touzeau:2025lwu}. Let us start by noting that the nulling procedure can be built from a set of arbitrary small bins. Physical degeneracy can be searched for in this limit. The bin's width, noted here $\Delta z$, can then be used as infinitesimal quantities. Starting from the expression of the two moments defined in Eq.~\eqref{eq:moments}, to first order in $\Delta z$ the coefficients $p_{i-2,i}$ is expressed as
\begin{equation}\label{Eq:degeneracy}
    p_{i-2,i}=1+\frac{\dd^2\xi_K/\dd z^2}{\dd\xi_K/\dd z}\Delta z-\frac{1}{n(z)} \frac{\dd n(z)}{\dd z}\Delta z,
\end{equation}
where $\xi_K(z)=1/F_K(\chi(z))$. We can easily identify two sources of invariance of the nulling coefficients with respect to parameterization: one is the dependence on the redshift distribution of galaxies that we do not consider here as it is assumed to be known; what then remains is the first term ($\xi_K''/\xi_K'$) which expresses the dependence on cosmological parameters, that is, the only functional of  $\xi_K(z)$ that $p_{ai}$ can constrain. This leads to some model degeneracy in the angular distance-redshift relation (for fixed ${\xi_K''}/{\xi_K'})$.

In practice, spanning a parameter space such as the matter density, the curvature, the dark energy equation of state, and its first derivative, $\{\Omega_{\rm m},\Omega_K,{\rm w}_0,{\rm w}_a\}$, can always be done. However, the degeneracy identified above can be approached in some redshift bins. This becomes apparent in the exploration of the $\{\Omega_{\rm m},{\rm w}_0\}$ parameter space we present in this Letter.

\end{document}